# Medical Image Segmentation Based on Deep Active Contour and Mean Curvature Loss Function

Xiao-qiang Zhai[1*], Zhi-feng Pang[1,2], Peng Zheng[2], Ze-wen Li[3], Yan-zhe Hou[4]

1 Henan Provincial Center for Applied Mathematics, Henan University, Kaifeng 475004, China

2 College of Mathematics and Statistics, Henan University, Kaifeng 475004, China

3 Advanced Institute of Finance, Henan University, Kaifeng 475004, China

4 School of Software, Henan Agricultural University, Zhengzhou 450000, China

** Correspondence: zhaixiaoqiang19@mails.ucas.ac.cn*

## Abstract

**Background and Objective:** Pixel-level supervision in medical image segmentation ignores the geometric priors of the target region. Chan-Vese losses restore region-level priors and boundary regularity but do not characterise local boundary geometry. We introduce a second-order geometric constraint that is formally defined, cheap and usable as a plug-and-play training objective.

**Methods:** We propose the Deep Active Contour and Mean Curvature (DACMC) loss, which embeds the mean curvature of the predicted boundary in the Chan-Vese region functional. A fixed fourth-order-accurate convolution kernel evaluates curvature without explicit differentiation. The loss adds no trainable parameters, no inference cost and 4.8% training time, with one hyper-parameter $\lambda = 10^{-3}$ shared by all settings. We evaluate DACMC on liver CT, spleen MRI and ISIC dermoscopy images with U-Net and CE-Net backbones, four metrics – Dice similarity coefficient (DSC), 95th-percentile Hausdorff distance (HD95), Jaccard similarity (JS) and average surface distance (ASD) – and four competing losses under an identical protocol, plus a five-point λ sweep, a wall-clock cost measurement and a five-seed replication with paired Wilcoxon signed-rank tests.

**Results:** DACMC ranks first or second in DSC in five of the six dataset–backbone settings, most consistently in distance metrics. On spleen MRI it reduces HD95 and ASD to 16.28 and 1.90 mm, and on ISIC dermoscopy it reduces HD95 to 7.08 mm. DSC varies by 0.0023 across five $\lambda$ values spanning three orders of magnitude; degradation appears only when curvature dominates. Five-seed replication shows that curvature stabilises training and removes occasional severe failures; mean differences from the closest competing loss do not reach statistical significance at five seeds.

**Conclusions:** DACMC is a formal, low-overhead computing method imposing second-order geometric regularity on segmentation boundaries, with the clearest benefit for small or scale-varying targets. Its single weight is insensitive over a wide interval and its cost is confined to training. Limitations include dataset size and partitioning, classical backbones and losses, and restriction to 2-D single-organ images.

# 1. Introduction

Medical image segmentation is a computer vision task whose aim is to partition an image into meaningful regions for further clinical analysis, diagnosis, treatment planning and measurement of disease progression [1]. High-precision segmentation results can help doctors and other medical professionals to better understand or track the changes of the region of interest. Along with the rapid development of deep learning, many data-driven techniques such as the U-Net [2], the SegNet [3], the VGGNet [4], and the GoogLeNet [5] have been applied to various medical imaging modalities. However, these techniques usually require accurate delineation of structures and regions of interest in complex and noisy images. Producing such delineations requires expert knowledge and is time-consuming and costly, which limits the scalability and applicability of segmentation methods. In addition, data-driven techniques have difficulty in recognising the object boundary precisely. The ethical and legal constraints associated with the medical data to preserve the privacy and security of sensitive patient information make the data collection and annotation task more challenging.

Despite the recent progress of using the deep learning method for medical image segmentation, the practical application still faces challenges due to its poor interpretability and the requirement of large labelled datasets [6], [7]. To improve the interpretability of deep learning strategies, coupling the variational energy methods as network modules, some classical Deep Variational Segmentation Models (DVS) have been incorporated into segmentation networks for end-to-end learning. This mainly uses the fact that the variational energy methods have better mathematical interpretability and also can realise the continuous, closed and smooth representation of target boundary. The DVS can be divided into two categories. One approach is to apply spatial regularisation to the network response by representing it as the solution of a variational problem [8]. Such a regularised response improves the back-propagation process and the learnable parameters. In addition, such a regulariser is flexible enough to be combined with other spatial priors, for instance a compactness term [9]. In practice, deep learning methods rely on loss functions to measure the discrepancy between the ground truth and the prediction, and the network parameters are optimised by numerical methods to minimise this loss. Based on these facts, another approach is to formulate new loss functions to improve upon the classical loss functions such as the cross-entropy loss (CE) [10] and the Dice loss [11]. These classical loss functions have two main limitations: they are pixel-wise loss functions that measure the similarity between the ground truth image and the prediction image, but the prior information is not taken into account. In addition, these classical loss functions in image segmentation tasks only consider the information from each individual pixel instead of the relationship between them and often lead to unexpected results in some complicated scenarios. However, the medical images usually contain much contextual information rather than point samples [12].

To obtain more accurate and plausible results [13], some recent approaches have introduced complex geometrical structures such as shape constraints [14], the topology specifications [15], the boundary distance [16], or the piecewise-constant region priors [17] to develop more sophisticated loss functions. However, integrating the shape prior still remains one of the most commonly used strategies toward anatomically meaningful predictions. In particular, the exploration of synergies between fully convolutional networks (FCNs) and the Active Contour Models (ACMs) [18] has recently attracted research interest [19]. These synergies utilise the fact that ACMs mainly use an enclosed and smooth curve to signify the target boundary, which is usually accomplished by minimising the associated energy function through the standard numerical optimisation methods or the dynamic programming [12], [20]. By embedding ACMs into deep neural networks, several methods have been proposed to combine the advantages of both data-driven and model-driven approaches [19], [21], [22], [23], [24]. However, these methods still lack an effective characterisation of the local geometric properties of the segmented region, such as the mean curvature. To address this limitation, we propose to incorporate the mean curvature as a geometric constraint into the loss function.

- We integrate the active contour model with the mean curvature into a loss function, where the mean curvature as a geometric prior guarantees the connectivity and smoothness of the network prediction result.
- To reduce the computational cost, we evaluate the mean curvature through the discrete mean curvature filter, a fixed convolution kernel that requires no explicit differentiation of the image surface.
- The experimental results on three public datasets – liver CT, spleen MRI and ISIC dermoscopy – with the U-Net and CE-Net backbones show that DACMC attains the best or the second-best Dice score in five of the six dataset-backbone settings, and that its most consistent gains are obtained in the boundary metrics, where the mean curvature acts as a boundary regulariser.

The remainder of this paper is organised as follows. Section 2 provides the preliminary background on active contour models and related loss function designs in medical image segmentation. Section 3 provides the mathematical background on mean curvature and its applications in image processing. Section 4 presents the proposed DACMC loss function in detail. Section 5 describes the experimental setup, datasets, and implementation details. Section 6 reports the experimental results, including quantitative comparisons, ablation studies, and visual analysis. Section 7 discusses the limitations of the present study and Section 8 concludes the paper.

# 2. Preliminary

The U-Net [2] has become the most commonly used structure in medical image segmentation tasks due to the introduction of spatial information. Its encoder progressively downsamples the input feature maps, while the decoder upsamples them back to the image resolution, and skip connections concatenate the encoder features with the corresponding decoder feature maps to preserve spatial details. In the U-Net architecture, the loss function plays an important role. Here we review several loss functions that are closely related to the proposed method.

## 2.1. Cross-entropy loss and Dice loss

Cross-entropy (CE) is a commonly used loss function for binary segmentation problems. It measures the dissimilarity between the predicted probability of a class and the true class label. The CE loss function is defined as

$$\mathcal{L}_{CE}(v,p) = -(\mathrm{v}\log(p) + (1-v)\log(1-p)), \qquad (1)$$

where v is the true class label, either 0 or 1, and p is the predicted probability of the positive class. Dice loss [25] is used to evaluate the similarity between the predicted segmentation mask and the ground truth mask. The loss function is defined as

$$L_{DC}(\mathcal{P},\mathcal{G}) = 1 - \frac{2\mathcal{P} \otimes \mathcal{G}}{|\mathcal{P}| + |\mathcal{G}|}, \quad (2)$$

where $\mathcal{P}$ is the predicted segmentation mask, $\mathcal{G}$ is the ground truth segmentation mask, $\mathcal{P} \otimes \mathcal{G}$ is the number of pixels that are in the intersection of the predicted and ground truth masks, and $|\mathcal{P}|$ and $|\mathcal{G}|$ respectively denote the total number of pixels in the predicted and ground truth segmentation results.

## 2.2. Mumford-Shah loss

The Mumford-Shah (MS) model [26] has been the key topic in classical image segmentation. The main technical difficulty herein is that the functional is non-convex and is not the lower semicontinuous envelope of a more amenable functional, making it difficult to find smooth approximations and to solve the minimisation in practice. Assuming the image to be a piecewise-constant region, the simple MS model can be reformulated as the following optimisation problem

$$\min_{u(\mathrm{x})\in\{0,1\}} \int_\Omega |\nabla u(\mathrm{x})|\mathrm{dx} + \mu_1 \int_\Omega |I(\mathrm{x}) - c_1|^2 u(\mathrm{x})\mathrm{dx} + \mu_2 \int_\Omega |I(\mathrm{x}) - c_2|^2 \big(1 - u(\mathrm{x})\big)\mathrm{dx}, \quad (3)$$

where $\Omega$ denotes the image domain, $\mu_1$ and $\mu_2$ are the tuning parameters, $I$ denotes the observed image, $u: \Omega \to \mathbb{R}$ is the segmentation function, and $\int |\nabla u| \, dx$ denotes the boundary length of the segmentation region, with $|\nabla u(x)| := \sqrt{\left(\nabla_{x_1} u(x)\right)^2 + \left(\nabla_{x_2} u(x)\right)^2}$.

$$c_1 = \frac{\int_\Omega I(\mathrm{x})u(\mathrm{x})\mathrm{dx}}{\int_\Omega u(\mathrm{x})\mathrm{dx}} \text{ and } c_2 = \frac{\int_\Omega I(\mathrm{x})(1 - u(\mathrm{x}))\mathrm{dx}}{\int_\Omega (1 - u(\mathrm{x}))\mathrm{dx}}$$

denote the mean values inside and outside the segmentation region, respectively. Problem (3) is non-convex, and many numerical methods [12], [27] have been proposed to solve it by convex relaxation, with $u \in [0,1]$. Eq. (3) is the two-phase piecewise-constant form of the MS functional and coincides with the Chan-Vese (CV) model [28], whose region and length terms are adopted in Section 4.

Based on the observation that the softmax layer of deep neural networks has close similarity to the characteristic function in the MS model (3), the authors in [29] proposed an MS-driven segmentation scheme where the loss function is defined by

$$\mathcal{L}(\bar{u}_\theta, \bar{c}_1, \bar{c}_2) := \int_\Omega |\nabla \bar{u}_\theta(\mathrm{x})|\mathrm{dx} + \mu_1 \int_\Omega |I(\mathrm{x}) - \bar{c}_1|^2 \bar{u}_\theta(\mathrm{x})\mathrm{dx} + \mu_2 \int_\Omega |I(\mathrm{x}) - \bar{c}_2|^2 (1 - \bar{u}_\theta(\mathrm{x}))\mathrm{dx}, \quad (4)$$

where $u_\theta$ denotes the softmax function of the output layer and

$$\bar{c}_1 = \frac{\int_\Omega I(\mathrm{x})\bar{u}_\theta(\mathrm{x})\mathrm{dx}}{\int_\Omega \bar{u}_\theta(\mathrm{x})\mathrm{dx}} \text{ and } \bar{c}_2 = \frac{\int_\Omega I(x)(1 - \bar{u}_\theta(\mathrm{x}))\mathrm{dx}}{\int_\Omega (1 - \bar{u}_\theta(\mathrm{x}))\mathrm{dx}}.$$

To use the labelled data, the authors in [29] proposed a hybrid loss function as

$$\mathcal{L}_1 := \beta \mathcal{L}(\bar{u}_\theta, \bar{c}_1, \bar{c}_2) + \mathcal{L}_{CE}(v, p), \quad (5)$$

where $\beta$ is the parameter.

## 2.3. Active contour with an elastic loss function

The elastica energy first appeared in Euler's work in 1744 on modelling torsion-free thin elastic rods. Later, Mumford [30] reintroduced elastica into computer vision for measuring the quality of interpolating curves in disocclusion or inpainting. Subsequently, the classical elastica-based segmentation model reviewed in [31] is formulated as

$$\min_{\phi(\mathrm{x}), c_1, c_2} \int_\Omega \left[\alpha + \beta \mathrm{div}\left(\frac{\nabla\phi(\mathrm{x})}{|\nabla\phi(\mathrm{x})|}\right)^2\right] |\nabla\phi(\mathrm{x})|\mathrm{dx} + \int_\Omega |I(\mathrm{x}) - c_1|^2 H(\phi(\mathrm{x}))\mathrm{dx} + \int_\Omega |I(\mathrm{x}) - c_2|^2 (1 - H(\phi(\mathrm{x})))\mathrm{dx}, \quad (6)$$

where $\alpha$ and $\beta$ are the tuning parameters, and the Heaviside function $H(\cdot)$ needs to be smoothed in numerical implementation to overcome its non-differentiability.

To integrate Euler's Elastica model into a deep learning framework, Chen et al. in [31] introduced a new loss function called Active Contour with an Elastic (ACE) described as follows

$$\begin{aligned}\mathcal{L}_2 = \int_\Omega \left[\alpha + \beta \operatorname{div}\left(\frac{\nabla \tilde{u}_\theta(\mathrm{x})}{|\nabla \tilde{u}_\theta(\mathrm{x})|}\right)\right] |\nabla \tilde{u}_\theta| \mathrm{dx} \\ + \int_\Omega |I(\mathrm{x}) - c_1|^2 \tilde{u}_\theta(\mathrm{x}) \mathrm{dx} + \int_\Omega |I(\mathrm{x}) - c_2|^2 (1 - \tilde{u}_\theta(\mathrm{x})) \mathrm{dx}\end{aligned}, \quad (7)$$

where $\tilde{u}_\theta$ denotes the predicted segmentation results and $I(\mathrm{x})$ denotes the binary ground truth defined by

$$I(\mathrm{x}) = \begin{cases} 1, & \text{if } \mathrm{x} \in \mathcal{F} \\ 0, & \text{if } \mathrm{x} \in \mathcal{B} \end{cases} \quad (8)$$

Here $\mathcal{F}$ denotes the segmented foreground region and $\mathcal{B}$ denotes the segmented background. $c_1$ is the internal greyscale average of the evolving curve and $c_2$ is the external greyscale average of the evolving curve. In particular, we set $I(x) = 1$ on $\mathcal{F}$ and $I(x) = 0$ on $\mathcal{B}$, giving $c_1 = 1$ and $c_2 = 0$.

# 3. Mathematical background

This section provides the mathematical preliminaries underlying the proposed DACMC loss function. We first review the definition of the mean curvature and its role in image processing, and then introduce the discrete curvature formulation adopted in this work. The mean curvature, which describes the local curvature of a surface [32], serves as the geometric prior that motivates our approach.

## 3.1. Mean curvature

Definition 1: Let $u: \Omega \rightarrow \mathbb{R}$ be the segmented image defined on the domain $\Omega \subset \mathbb{R}^2$, and denote the corresponding surface by $x_3 = \tilde{u}(x_1, x_2)$. Then the mean curvature of this surface is defined by

$$H(\tilde{u}) = \text{div}\left(\frac{\nabla \tilde{u}}{\sqrt{1 + |\nabla \tilde{u}|^2}}\right) \quad (9)$$

Mean curvature-based models can reduce the staircase effect that often occurs in total variation-based models, while preserving object edges during denoising. These properties make the mean curvature an effective geometric prior for image processing tasks.

In the ACE loss function (Eq. (7)), the curvature term involves high-order derivatives, which complicates the development of efficient numerical algorithms. The non-smoothness of this term also increases the difficulty of back-propagation in deep learning networks. To address these issues, the discrete curvature adopted in this work is defined as

$$C = \sum_{i=1}^{m} \sum_{j=1}^{n} \left| \frac{u_{i,j}^* v_{i,j}}{v_{i,j}^* + \epsilon} \right| \quad (10)$$

## 3.2. Mean curvature in image processing

At any point on a surface, the normal curvature varies with the choice of the tangent direction. The maximum and minimum values of the normal curvature are called the principal curvatures [33], and the mean curvature H, whose second-order hyperbolic formulation arises naturally in geometric flows [34], is defined as the mean of the two principal curvatures: $H$

$$H = \frac{1}{2}(k_1 + k_2) \quad (11)$$

$$H(U) = \frac{\left(1 + U_y^2\right)U_{xx} - 2U_x U_y U_{xy} + \left(1 + U_x^2\right)U_{yy}}{2\left(1 + U_x^2 + U_y^2\right)^{3/2}}, \quad (12)$$

where $U$ denotes the target image, and the remaining terms ($U_x, U_y, Uxy, Uxx, Uyy$) are computed by central finite differences in the discrete implementation.

Zhu [32] employed mean curvature parametrisation as a denoising model with an $L^1$-norm regulariser.

$$E(\hat{u}) = \lambda \int_{\Omega} |H(\hat{u})| dx + \int_{\Omega} (f - \hat{u})^2 dx \quad (13)$$

The first term of Eq. (13) is the mean curvature regularisation term, and the second is the data fitting term. Because the $L^1$-norm regulariser does not penalise jumps in pixel intensity, this model maintains image contrast better than the Rudin-Osher-Fatemi (ROF) model [35]. Zhu et al. [36] showed that the mean curvature approach can preserve corner contrast, smooth object surfaces, and maintain greyscale intensity. The $L^1$-norm of mean curvature also preserves geometric characteristics of object shapes, particularly at corners, as shown below:

$$E_{MAMS}(\hat{u}) = \frac{1}{2}\int_{\Omega} (f - \hat{u})^2 dx + \frac{\mu}{2}\int_{\Omega} |\nabla\hat{u}|^2 dx + \lambda \int_{\Omega} \left|\nabla \cdot \left(\frac{\nabla\hat{u}}{\sqrt{|\nabla\hat{u}|+1}}\right)\right| dx \quad (14)$$

These properties motivate us to incorporate mean curvature as a geometric prior into the segmentation loss function, which we present in the next section.

# 4. Proposed method

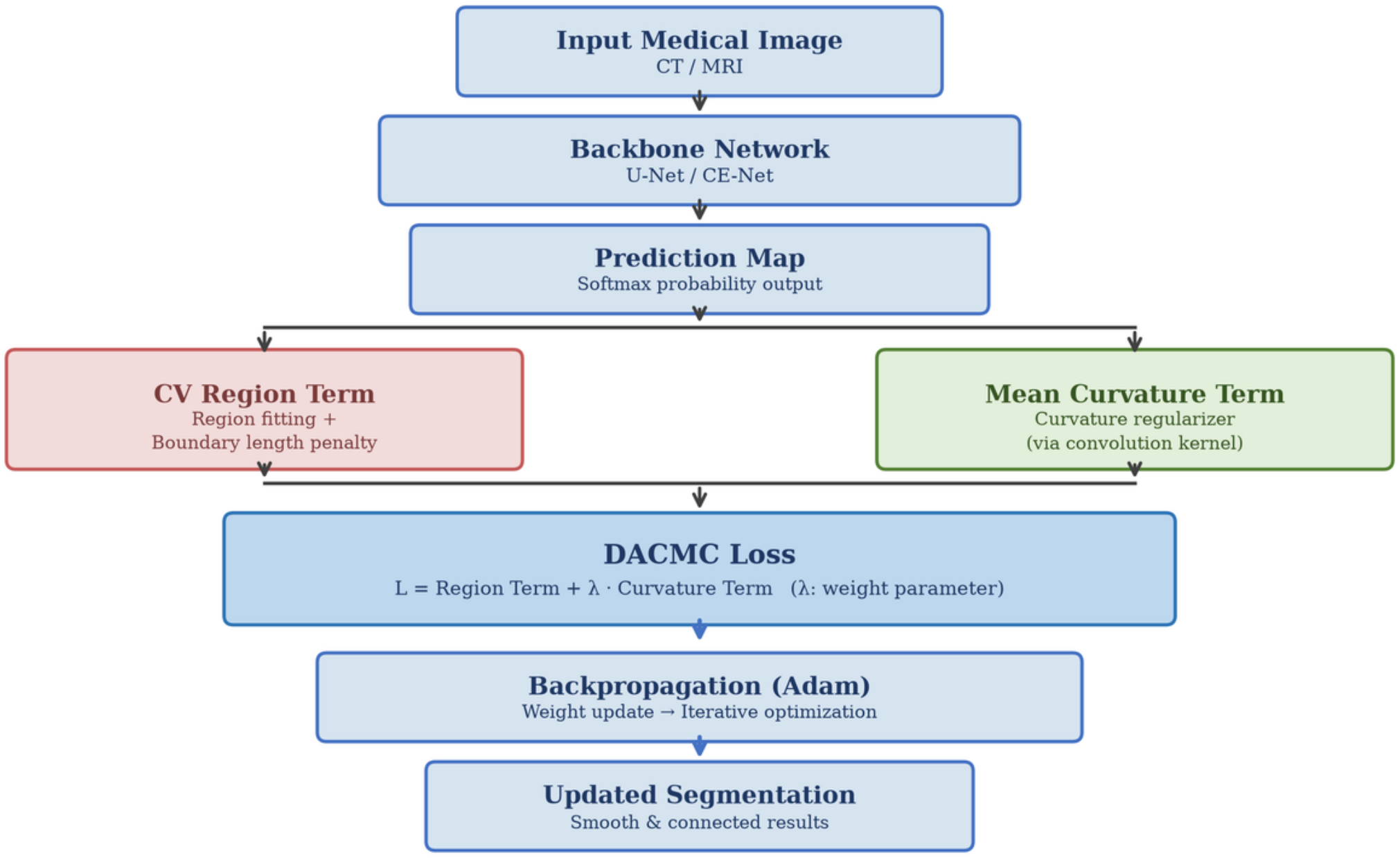

Figure 1. Overall training pipeline of the proposed DACMC loss function. A medical image is fed through a backbone network such as U-Net or CE-Net to produce a prediction map. The DACMC loss combines a CV-based region term and a mean-curvature term, evaluated through a fixed convolution kernel, to supervise training. The combined loss is back-propagated via Adam to update the network weights.

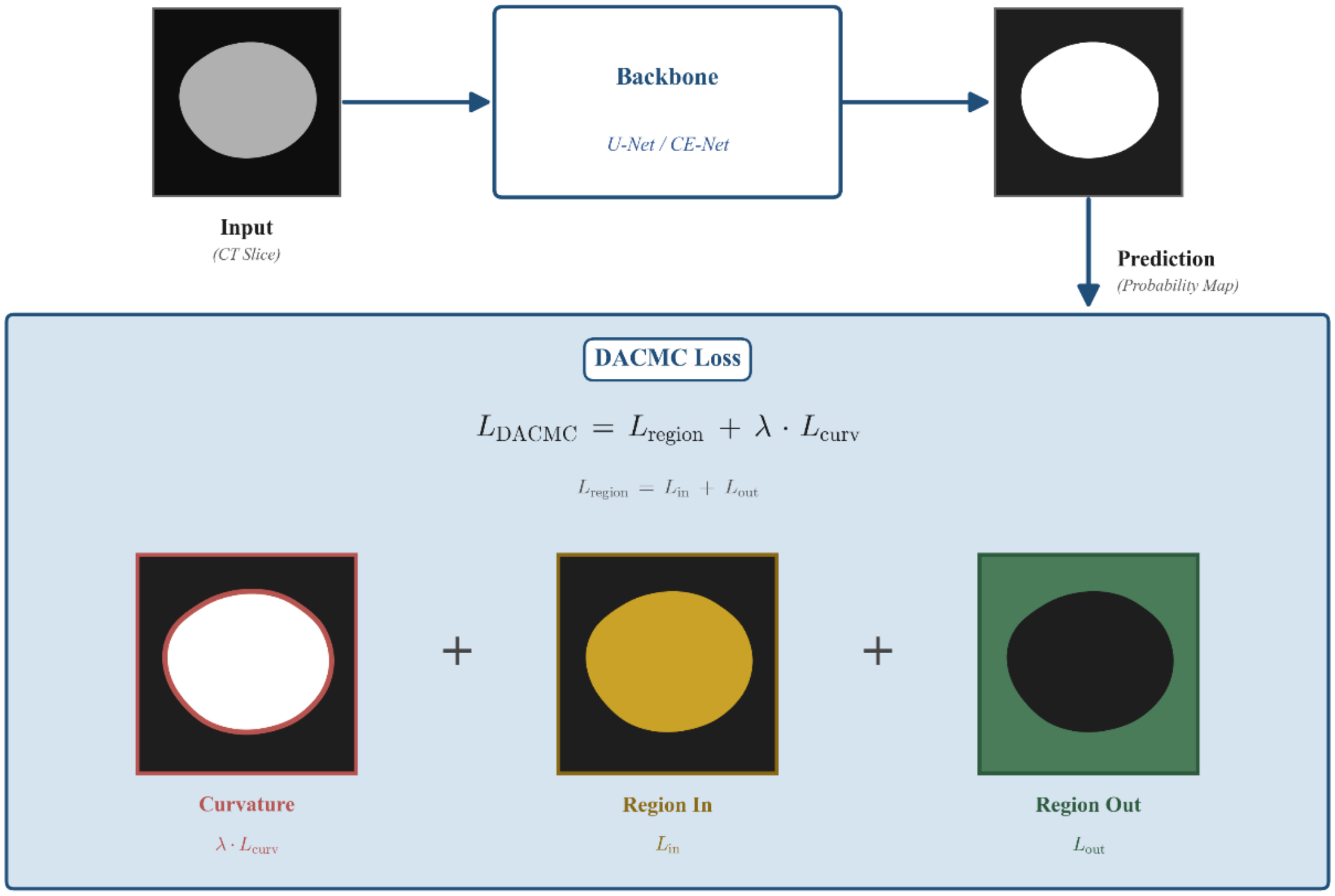

Figure 2. The general overview of the proposed DACMC loss, integrating curvature loss terms and region terms as geometric natural constraints for image segmentation.

The overall training pipeline is shown in Figure 1. In the classical Chan-Vese model [28], multiple iterations are performed to minimise the energy functional to find the best active contour, which is the boundary of the segmentation target. On this basis, as shown in Figure 2, we introduce the mean curvature term, because the internal voids and non-smoothness of the network prediction results will lead to an increase in the mean curvature. Using the mean curvature as a constraint can ensure the connectivity and boundary smoothness of the network prediction results. In the work of Xing et al. [37], a new curvature loss term is proposed, which can effectively preserve the boundary smoothness and regional connectivity of the segmented target. We also adopt this approach to improve the loss function. The proposed DACMC loss function is defined on the two-dimensional image as follows:

$$L_{DACMC} = \sum_{i=1}^{m} \sum_{j=1}^{n} \left( u_{i,j}\left(c_1 - v_{i,j}\right)^2 + \left(1 - u_{i,j}\right)\left(c_2 - v_{i,j}\right)^2 \right) + \lambda \mathcal{C}, \quad (15)$$

where the curvature $C$ is defined by Eq. (10), and λ is the tuning parameter that balances the region term and the curvature term.

In practice, the proposed curvature term evaluates the curvature response of u and v, respectively, using the discrete mean curvature filter of Gong and Sbalzarini [38], which is derived from Euler's theorem and avoids explicit computation of the image surface derivatives, thereby reducing the computational cost on the GPU during training. Section 6.5 quantifies the relation between this filter and the analytic mean curvature, and the cost that the term adds.

The filter is evaluated as follows, where ∗ denotes the convolution operator:

$$H(u) \approx \text{Kernel} * u \quad (16)$$

$$\text{Kernel} = \begin{bmatrix} -\frac{1}{16} & \frac{5}{16} & -\frac{1}{16} \\ \frac{5}{16} & -1 & \frac{5}{16} \\ -\frac{1}{16} & \frac{5}{16} & -\frac{1}{16} \end{bmatrix} \quad (17)$$

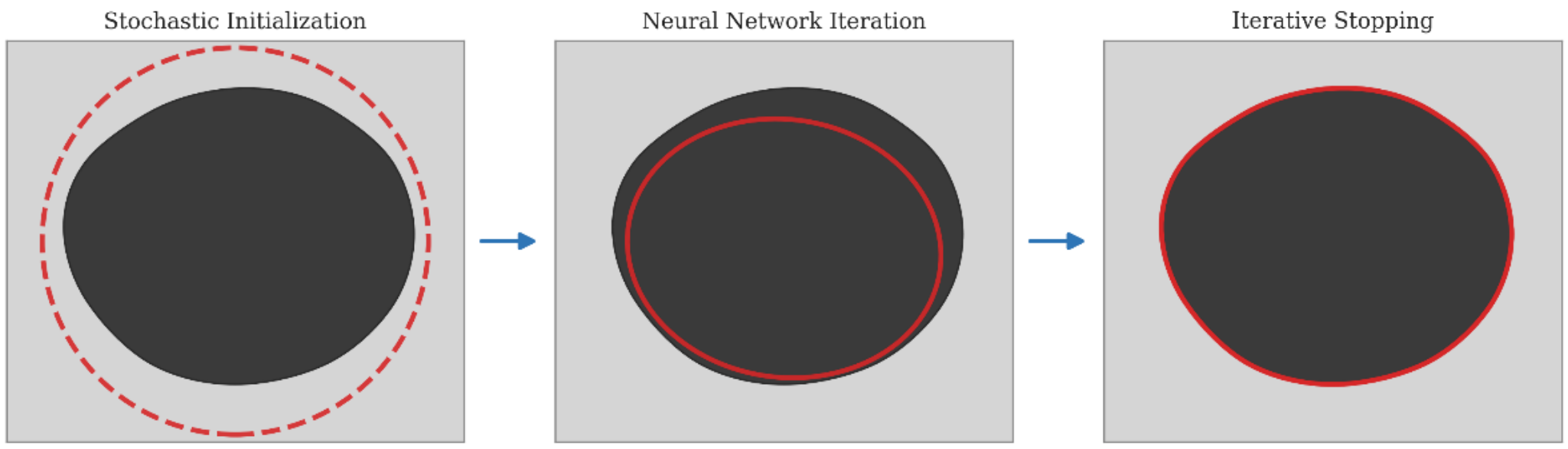

Figure 3. Illustration of the iterative evolution of the segmentation boundary predicted by the network. The black and grey regions denote the foreground and background, respectively, and the red curve denotes the predicted segmentation boundary. The region term of the DACMC loss is minimised only when the predicted boundary coincides with the object boundary.

where, in the CV model, $\mathbf{c_1}$ and $\mathbf{c_2}$ are the spatial averages of the image intensity u inside and outside the contour, respectively, and are updated at each iteration as defined in Eq. (18):

$$\begin{cases} c_1 = \frac{\int v \cdot u dx}{\int u dx} \\ c_2 = \frac{\int v(1-u) dx}{\int (1-u) dx} \end{cases} \quad (18)$$

In Figure 3, the black and grey sections represent the foreground and background, respectively, and the red curve corresponds to the segmentation boundary predicted by the neural network. The network is trained based on the region terms of Eq. (15), and the predicted segmentation boundary then starts to evolve. $c_1$ and $c_2$ represent the pixel averages inside and outside the segmentation boundary after each iteration, as in Eq. (18). Upon reaching the stopping condition for iteration, $c_1$ and $c_2$ closely align with the foreground and background, respectively.

In the supervised learning framework, the method also fixes $c_1$ and $c_2$ as foreground and background, simply defined as constants with a positive number added to ensure the stability of the computation. Overall, the proposed DACMC loss function is designed, integrating the region term and the mean curvature term to ensure the accuracy of the segmentation.

# 5. Experimental setup

## 5.1. Dataset

To evaluate the effectiveness of the designed loss function, three publicly available datasets were used, covering two imaging modalities and a range of target sizes. (i) A two-dimensional liver CT dataset, available at https://www.kaggle.com/datasets/zxcv2022/digital-medical-images-for--download-resource, which contains 100 contrast-enhanced abdominal CT slices with corresponding liver masks; the slices were partitioned by stratifying on the liver area, so that the training and test subsets cover the same range of target sizes. (ii) The spleen MRI subset of the CHAOS challenge [39], from which 165 axial slices containing a visible spleen were selected from 20 subjects; the data were partitioned at the subject level, the 4 subjects reserved for testing being absent from the training data, and the remaining 16 subjects supplying the training and validation slices. (iii) A publicly available dermoscopy dataset derived from the ISIC archive, from which 500 images with binary lesion masks were used, again partitioned by stratifying on the lesion area. All images and masks were resized to 512 × 384 pixels. The resulting splits are 69/15/16 slices for liver

CT, 120/20/25 slices for spleen MRI and 360/40/98 images for ISIC, where the three numbers denote the training, validation and test subsets, respectively.

## 5.2. Network architecture

In this study, we utilised the U-Net [2] architecture as a framework for image segmentation. U-Net is an end-to-end encoder-decoder network that incorporates skip connections between the upsampling and downsampling paths, so that high-resolution features are preserved and segmentation outputs can be produced efficiently. Its simplicity and modest requirement for labelled samples make it well suited to medical image segmentation, where annotations are costly to obtain. In addition to U-Net, we employed the CE-Net [40] architecture. Between its encoder and decoder, a context extractor combines a dense atrous convolution (DAC) block and a residual multi-kernel pooling (RMP) module to capture multi-scale context and generate higher-level semantic feature maps.

## 5.3. Training process

All training was performed in Python and PyTorch with CUDA, using an Intel Xeon 5218R CPU and an NVIDIA RTX 5000 GPU. The batch size is set to 4 and the number of epochs is 300, and the models were trained using the Adam optimiser. To compare these loss functions in a reasonable way, we used a learning rate of 0.0001 for all loss functions. No data augmentation or post-processing techniques were applied, so that all loss functions are compared under identical training and evaluation conditions. Because the curvature term enters only the training objective, the proposed loss adds no parameters to the network and no computation to the forward pass, so that its inference cost is identical to that of the baseline losses.

The checkpoint with the highest Dice score on the validation subset was retained for evaluation on the test set, and the test sets were never used for model selection. The validation subsets were carved out of the training partitions with a fixed random split and seed 42: 15 of 84 liver slices, 20 of 140 spleen slices and 40 of 400 dermoscopy images were held out. Since these subsets come from the same partition as the training data, a note of caution applies when they are used for hyper-parameter selection; this is discussed in the Conclusion.

DACMC involves a single hyper-parameter, the curvature weight $\lambda$ in Eq. (15). $\lambda$ was selected on the validation subsets alone. On spleen MRI and ISIC the sweep covered $\{10^{-4}, 5\times10^{-4}, 10^{-3}, 5\times10^{-3}\}$ (Table 3), and on liver CT over five values in $[0, 10^{-3}]$ (Table 4). The single value $\lambda = 10^{-3}$ was adopted for every dataset and every backbone. Using one common value rather than a separately tuned $\lambda$ per configuration is the more conservative choice, since it rules out any form of per-dataset tuning. $\lambda = 10^{-3}$ was the validation optimum in three of the four dataset-backbone configurations in which the sweep was repeated, and on liver CT the response over the whole interval $[0, 10^{-3}]$ is a flat plateau. The test DSC varies by only 0.0023 across five values of $\lambda$ spanning three orders of magnitude, so the results are insensitive to the exact choice of $\lambda$ below the collapse point and the reported performance is not an artefact of fine-grained validation tuning. The competing losses were used with their original settings: CE and Dice have no hyper-parameters, the active contour loss (AC) uses its published weighting ($\beta = 1$), and ACE uses the $(\alpha, \beta) = (10^{-4}, 1)$ reported in its original formulation. For completeness, ACE was additionally trained with $\alpha = 10^{-3}$, the value adopted for DACMC, in all six configurations; its performance was lower in 11 of the 12 comparisons and never changed the ranking of the five losses in Tables 1 and 2. With $\alpha = 10^{-3}$, the DSC of ACE on spleen MRI decreases from 0.8303 to 0.8228 with U-Net and from 0.7984 to 0.7823 with CE-Net, and its HD95 on ISIC increases from 10.44 to 13.26 mm with CE-Net. The reported comparison is therefore not affected by the choice of the ACE weight.

## 5.4. Metrics

Four metrics are used to evaluate segmentation performance, namely the Dice similarity coefficient (DSC), the Jaccard similarity (JS, also known as the intersection over union), the 95th-percentile

Hausdorff distance (HD95) and the average surface distance (ASD). The DSC and the JS both measure the overlap between the predicted segmentation and the ground truth, and are 0 for disjoint regions and 1 for a perfect match. The HD95 is the distance threshold exceeded by only 5% of the boundary point-to-point distances, and the ASD is the mean of these distances. The HD95 is therefore insensitive to a small fraction of gross boundary errors, whereas the ASD is sensitive to the alignment of the two boundaries as a whole.

# 6. Experiments

## 6.1. Comparison to other loss functions

The widely used CE, Dice, active contour (AC) and ACE loss functions are compared with the proposed DACMC loss function. Two standard convolutional neural networks (CNNs) for two-dimensional medical image segmentation, U-Net and CE-Net, serve as backbones to evaluate the performance of the DACMC loss function.

**Table 1.** Quantitative analysis results of U-Net on the liver CT, spleen MRI and ISIC dermoscopy datasets (mean ± standard deviation).

| Loss | DSC ↑ | HD95 ↓ | JS ↑ | ASD ↓ |
|---|---|---|---|---|
| **Liver CT** | | | | |
| CE | 0.9824 ± 0.0048 | **0.00 ± 0.00** | 0.9655 ± 0.0092 | 0.03 ± 0.04 |
| Dice | 0.9818 ± 0.0084 | **0.00 ± 0.00** | 0.9643 ± 0.0158 | 0.03 ± 0.04 |
| AC | 0.9833 ± 0.0055 | **0.00 ± 0.00** | 0.9671 ± 0.0104 | **0.02 ± 0.04** |
| ACE | 0.9817 ± 0.0050 | **0.00 ± 0.00** | 0.9641 ± 0.0095 | 0.03 ± 0.03 |
| DACMC | **0.9838 ± 0.0040** | **0.00 ± 0.00** | **0.9681 ± 0.0075** | 0.03 ± 0.03 |
| **Spleen MRI** | | | | |
| CE | 0.8139 ± 0.1439 | 29.34 ± 35.93 | 0.7082 ± 0.1833 | 7.67 ± 14.17 |
| Dice | 0.8139 ± 0.1738 | **21.12 ± 32.57** | 0.7163 ± 0.2072 | 5.59 ± 13.92 |
| AC | 0.8019 ± 0.1263 | 26.43 ± 35.35 | 0.6861 ± 0.1606 | 6.06 ± 13.32 |
| ACE | 0.8303 ± 0.1103 | 25.82 ± 34.21 | 0.7240 ± 0.1502 | 5.54 ± 9.77 |
| DACMC | **0.8348 ± 0.1102** | 26.51 ± 34.26 | **0.7307 ± 0.1503** | **5.47 ± 9.18** |
| **ISIC dermoscopy** | | | | |
| CE | 0.8500 ± 0.1641 | 17.26 ± 29.98 | 0.7657 ± 0.1900 | 3.42 ± 12.03 |
| Dice | 0.8674 ± 0.1532 | **13.95 ± 26.42** | 0.7891 ± 0.1753 | 3.08 ± 9.42 |
| AC | **0.8778 ± 0.1239** | 14.92 ± 29.45 | **0.7989 ± 0.1542** | 2.92 ± 9.75 |
| ACE | 0.8574 ± 0.1566 | 15.95 ± 31.73 | 0.7748 ± 0.1818 | 3.43 ± 11.55 |

| Loss | DSC ↑ | HD95 ↓ | JS ↑ | ASD ↓ |
|---|---|---|---|---|
| DACMC | 0.8633 ± 0.1485 | 14.80 ± 26.24 | 0.7816 ± 0.1735 | **2.28 ± 7.95** |

*DSC: Dice similarity coefficient; HD95: 95th percentile Hausdorff distance (mm); JS: Jaccard similarity; ASD: average surface distance (mm). The best value in each column is shown in bold.*

*All metrics are reported as mean ± standard deviation over the test set. DACMC uses λ = $10^{-3}$ throughout.*

**Table 2.** Quantitative analysis results of CE-Net on the liver CT, spleen MRI and ISIC dermoscopy datasets (mean ± standard deviation).

| Loss | DSC ↑ | HD95 ↓ | JS ↑ | ASD ↓ |
|---|---|---|---|---|
| **Liver CT** | | | | |
| CE | 0.9772 ± 0.0097 | **0.00 ± 0.00** | 0.9557 ± 0.0184 | 0.09 ± 0.20 |
| Dice | 0.9775 ± 0.0078 | **0.00 ± 0.00** | 0.9561 ± 0.0149 | 0.05 ± 0.05 |
| AC | **0.9799 ± 0.0074** | **0.00 ± 0.00** | **0.9607 ± 0.0141** | **0.04 ± 0.06** |
| ACE | 0.9775 ± 0.0073 | **0.00 ± 0.00** | 0.9562 ± 0.0139 | 0.05 ± 0.09 |
| DACMC | 0.9780 ± 0.0098 | 0.06 ± 0.24 | 0.9571 ± 0.0185 | 0.05 ± 0.06 |
| **Spleen MRI** | | | | |
| CE | 0.7723 ± 0.1504 | 27.14 ± 28.98 | 0.6515 ± 0.1842 | 8.19 ± 10.80 |
| Dice | **0.8283 ± 0.0837** | 19.06 ± 23.15 | **0.7153 ± 0.1170** | 4.04 ± 6.17 |
| AC | 0.7775 ± 0.1639 | 22.45 ± 23.22 | 0.6616 ± 0.1929 | 6.50 ± 9.91 |
| ACE | 0.7984 ± 0.1370 | 18.01 ± 21.35 | 0.6841 ± 0.1726 | 3.74 ± 7.26 |
| DACMC | 0.8223 ± 0.1180 | **16.28 ± 20.04** | 0.7145 ± 0.1636 | **1.90 ± 3.79** |
| **ISIC dermoscopy** | | | | |
| CE | 0.8910 ± 0.1083 | 11.55 ± 25.75 | 0.8169 ± 0.1403 | 2.29 ± 7.06 |
| Dice | 0.8792 ± 0.1560 | 9.95 ± 23.67 | 0.8089 ± 0.1779 | 1.41 ± 5.42 |
| AC | **0.8984 ± 0.0979** | 12.64 ± 31.39 | 0.8271 ± 0.1321 | 3.00 ± 9.61 |
| ACE | 0.8879 ± 0.1247 | 10.44 ± 25.33 | 0.8153 ± 0.1537 | 2.60 ± 8.71 |
| DACMC | 0.8944 ± 0.1407 | **7.08 ± 14.67** | **0.8284 ± 0.1560** | **1.14 ± 3.20** |

*DSC: Dice similarity coefficient; HD95: 95th percentile Hausdorff distance (mm); JS: Jaccard similarity; ASD: average surface distance (mm). The best value in each column is shown in bold.*

*All metrics are reported as mean ± standard deviation over the test set. DACMC uses λ = $10^{-3}$ throughout.*

**Table 3.** Sensitivity of DACMC to its curvature weight λ, measured by test DSC.

| Dataset / Backbone | λ = $10^{-4}$ | λ = $5\times10^{-4}$ | λ = $10^{-3}$ | λ = $5\times10^{-3}$ |
|---|---|---|---|---|
| Liver CT / U-Net | 0.9830 | — | **0.9838** | — |
| Liver CT / CE-Net | 0.9799 | 0.9784 | **0.9776** | — |
| Spleen MRI / U-Net | 0.8164 | 0.8245 | **0.8348** | 0.7968 |
| Spleen MRI / CE-Net | 0.7800 | 0.7981 | **0.8223** | collapse |
| ISIC dermoscopy / U-Net | 0.8586 | 0.8567 | **0.8633** | 0.8645 |

| Dataset / Backbone | λ = $10^{-4}$ | λ = $5\times10^{-4}$ | λ = $10^{-3}$ | λ = $5\times10^{-3}$ |
|---|---|---|---|---|
| ISIC dermoscopy / CE-Net | 0.8840 | 0.8927 | **0.8944** | 0.8821 |

*"—" indicates values that were not run. On liver CT the sweep was performed over five values in [0, $10^{-3}$] instead (Table 4). "collapse" denotes a fully degenerate prediction with empty masks and DSC = 0. $\lambda = 10^{-3}$ is the value used in Tables 1 and 2. Values for liver CT / CE-Net are taken from that dedicated sweep; the main-experiment runs in Tables 1 and 2 give 0.9780 at $\lambda = 10^{-3}$, a difference of less than 0.0005 DSC that is well within one standard deviation (0.0098) and arises from independent training runs.*

**Table 4.** Full five-point λ sweep of DACMC on liver CT with CE-Net.

| λ | DSC ↑ | HD95 ↓ | JS ↑ | ASD ↓ |
|---|---|---|---|---|
| 0 | 0.9776 ± 0.0092 | 0.00 | 0.9564 | 0.14 |
| $5\times10^{-5}$ | 0.9787 ± 0.0073 | 0.00 | 0.9584 | 0.05 |
| $10^{-4}$ | 0.9799 ± 0.0059 | 0.00 | 0.9607 | 0.05 |
| $5\times10^{-4}$ | 0.9784 ± 0.0087 | 0.00 | 0.9579 | 0.08 |
| $10^{-3}$ | 0.9776 ± 0.0096 | 0.00 | 0.9563 | 0.05 |

*The curve is a flat plateau for $\lambda \le 10^{-3}$, so the choice of λ within this range is not critical.*

# 6.2. Performance

Tables 1 and 2 report the quantitative results of U-Net and CE-Net on the three datasets, and the best value in each column of both tables is highlighted in bold. On the liver CT dataset all five losses are close to saturation and the differences among them are small. With U-Net, DACMC attains the highest DSC of 0.9838 ± 0.0040 and the highest JS of 0.9681 ± 0.0075; with CE-Net, the active contour loss AC is marginally ahead in DSC, at 0.9799 ± 0.0074 against 0.9780 ± 0.0098 for DACMC. The gap of 0.002 is smaller than the standard deviation of any single method on this dataset, and all five losses reach HD95 ≈ 0 mm, so that this metric no longer discriminates between them. The liver is a large, high-contrast organ whose boundary coincides with a strong intensity edge, which leaves little room for a geometric regulariser to contribute. On the spleen MRI dataset, where the spleen varies strongly in size across slices and its boundary is frequently partially obscured by surrounding tissue, the benefit of the curvature term becomes apparent. With U-Net, DACMC achieves the best DSC of 0.8348 ± 0.1102, ahead of the ACE baseline at 0.8303 ± 0.1103, the best JS of 0.7307 ± 0.1503 and the lowest ASD of 5.47 mm. With CE-Net, DACMC obtains the lowest HD95, 16.28 mm against 18.01 mm for ACE and 19.06 mm for Dice, and the lowest

ASD of 1.90 mm, a 49% reduction relative to ACE, while its DSC and JS of 0.8223 and 0.7145 rank second, behind the Dice loss on these two overlap measures. On the ISIC dermoscopy dataset the pattern is again dominated by the boundary metrics. With CE-Net, DACMC gives the lowest HD95, 7.08 mm, which is 29% below the next best value of 9.95 mm obtained with Dice, as well as the highest JS of $0.8284 \pm 0.1560$ and the lowest ASD of 1.14 mm, while AC reaches the highest DSC, 0.8984 against 0.8944 for DACMC. The U-Net results are less favourable. DACMC gives the lowest ASD of 2.28 mm but ranks third in DSC, behind AC at 0.8778 and Dice at 0.8674, with its own DSC at 0.8633. These results support a specific rather than a generic claim. The mean curvature term acts as a boundary regulariser. Its benefit is concentrated in the distance-based metrics, where DACMC ranks first or second in 10 of the 12 metric-configuration combinations, and it is most visible on the two datasets whose targets are small and vary in scale. It does not systematically improve region-overlap metrics on datasets where those are already saturated; on liver CT the five losses are statistically indistinguishable. Across the six dataset-backbone settings, DACMC is the best or the second-best method in DSC in five, achieves the best ASD in four and the best JS in three. Figures 4–6 illustrate the corresponding segmentation results: on liver CT all methods are visually close, as the quantitative overlap suggests; on spleen MRI, AC and ACE occasionally lose internal connectivity in slices where the target changes abruptly in scale, whereas DACMC preserves a closed and smooth contour; on the dermoscopy images, the advantage appears at the lesion margin, where the distance-based metrics improve most.

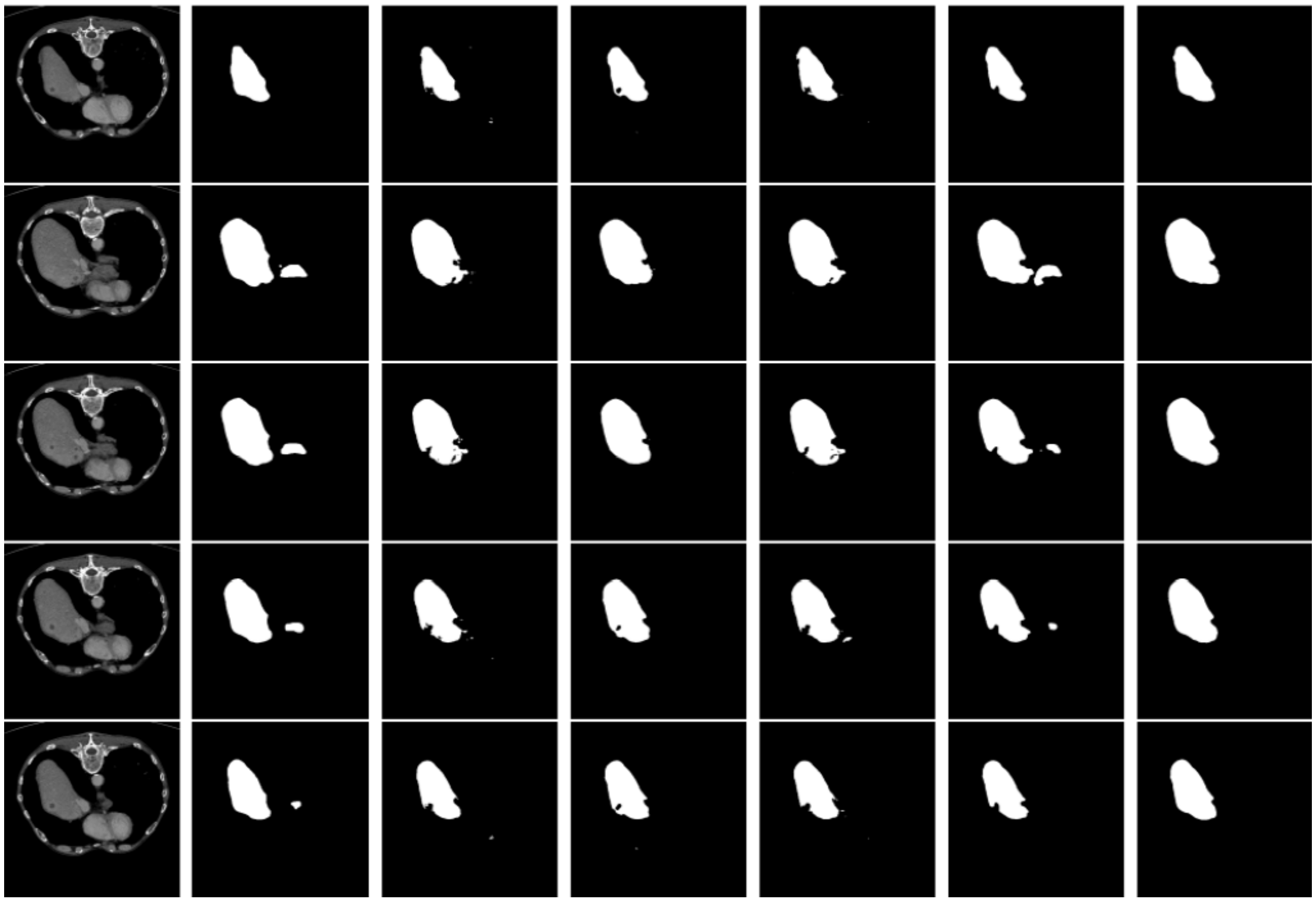

Figure 4. Segmentation results on the liver CT dataset using U-Net. From left to right: original image, Ground Truth, CE, Dice, AC, ACE, and DACMC.

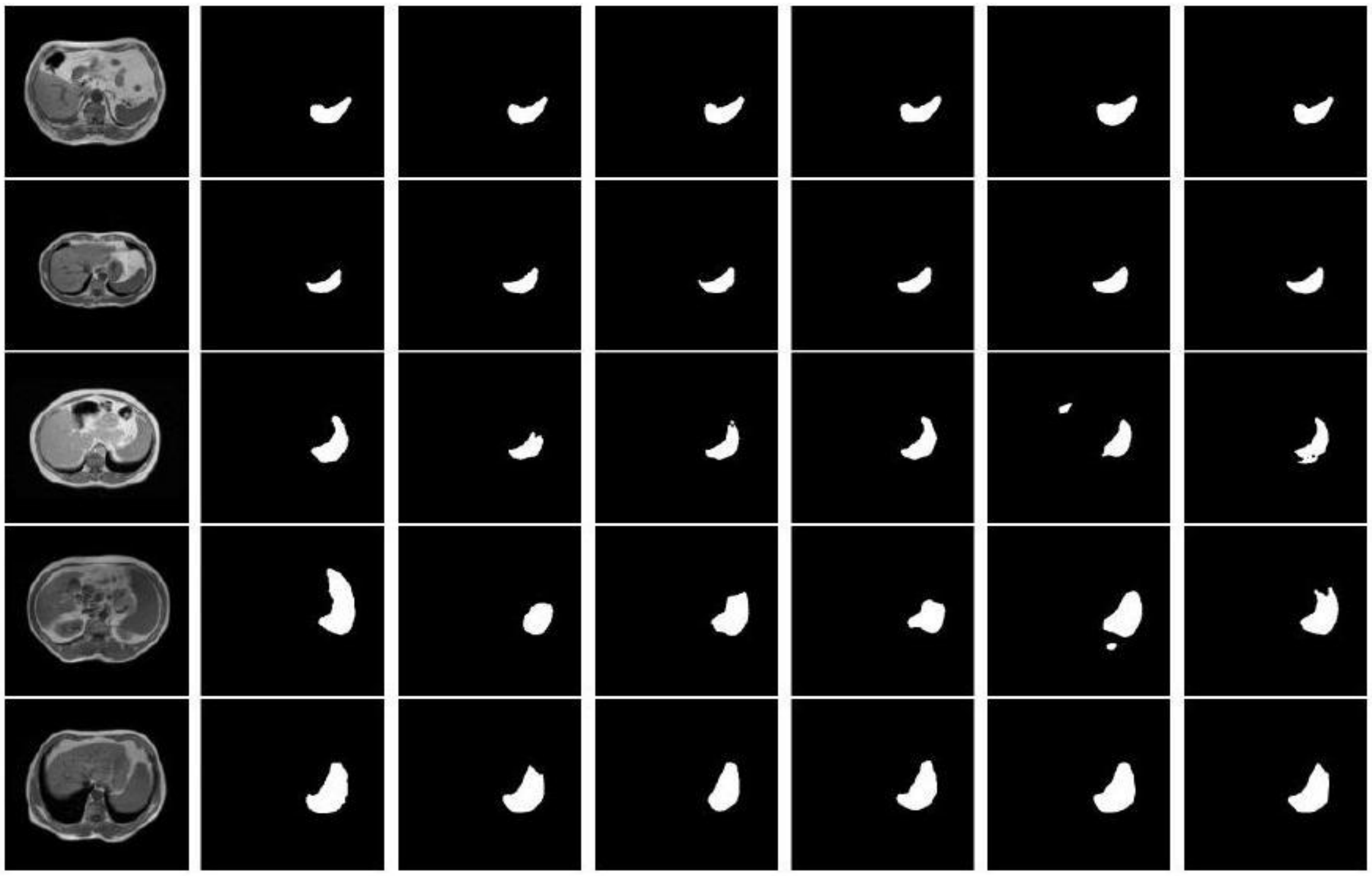

Figure 5. Segmentation results on the spleen MRI dataset using U-Net. From left to right: original image, Ground Truth, CE, Dice, AC, ACE, and DACMC.

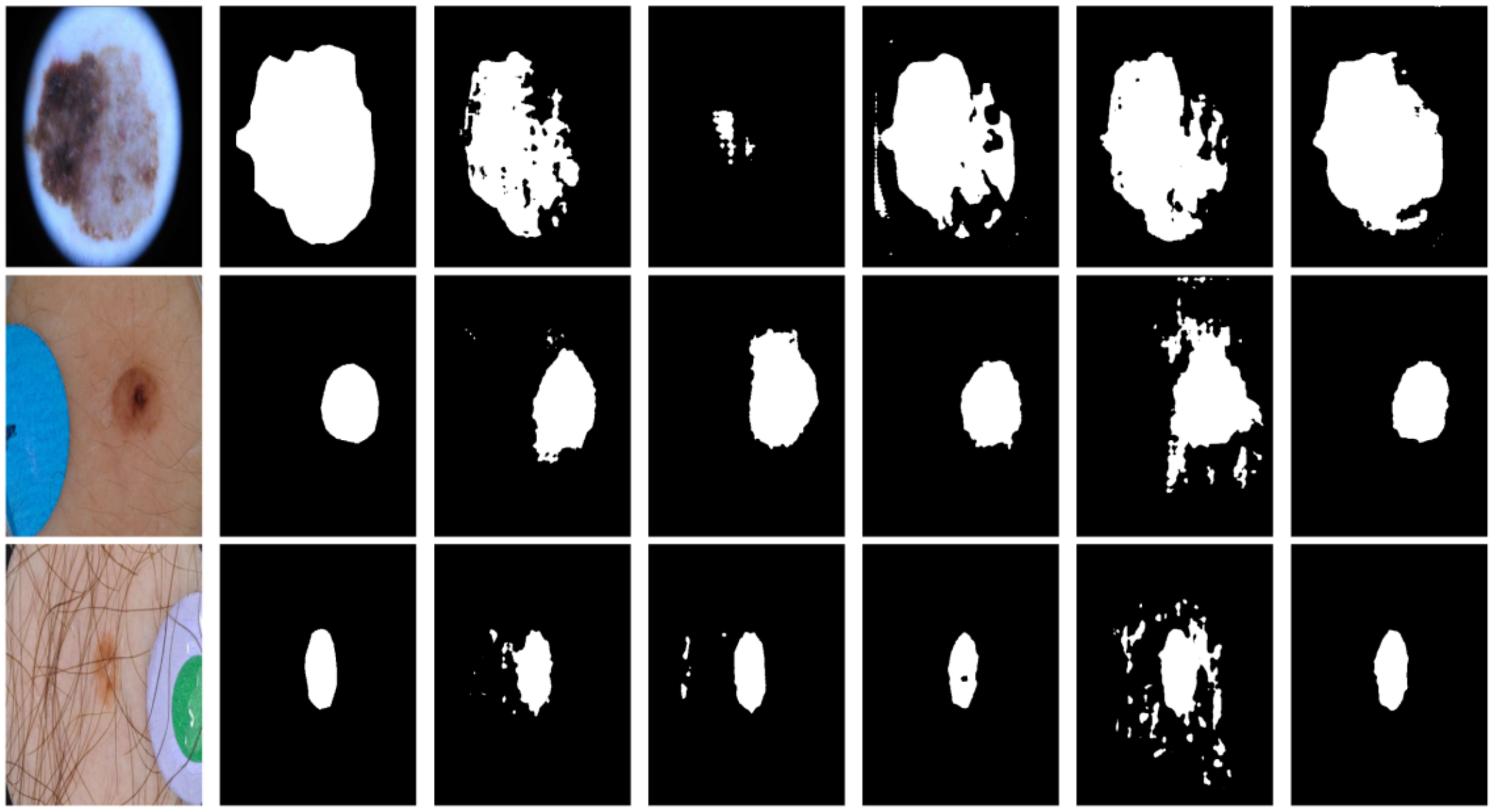

Figure 6. Segmentation results on the ISIC dermoscopy dataset using CE-Net. From left to right: original image, Ground Truth, CE, Dice, AC, ACE, and DACMC.

## 6.3. Ablation study

The ablation study examines the progressive contribution of each component of the objective by comparing CE, Dice, AC, ACE and DACMC under an identical protocol, as reported in Tables 1 and 2. On liver CT with U-Net, replacing the cross-entropy baseline of 0.9824 ± 0.0048 with the Dice loss leaves the DSC essentially unchanged at 0.9818 ± 0.0084; adding the active contour region term through AC gives 0.9833 ± 0.0055, and the elastica term of ACE gives 0.9817 ± 0.0050. DACMC, which replaces the elastica energy with an explicit mean curvature regulariser, attains the best DSC of 0.9838 ± 0.0040 and the best JS of 0.9681 ± 0.0075. All five losses reach HD95 ≈ 0 mm on this dataset, so the contribution of the geometric terms cannot be resolved by boundary metrics here. The separation becomes clear on the two harder datasets. On spleen MRI with U-Net, AC, at 0.8019, trails the cross-entropy and Dice losses, which are essentially tied at 0.8139, whereas the two geometry-aware losses lead. ACE reaches 0.8303 and DACMC reaches 0.8348, the best value on this dataset, together with the best JS of 0.7307 and the lowest ASD of 5.47 mm. On spleen MRI with CE-Net the same contrast appears in the boundary metrics. DACMC gives an HD95 of 16.28 mm and an ASD of 1.90 mm against 18.01 mm and 3.74 mm for ACE, while the Dice loss remains slightly better in DSC, at 0.8283 against 0.8223. On ISIC with CE-Net, DACMC reduces HD95 to 7.08 mm compared with 10.44 mm for ACE and 9.95 mm for Dice, and reaches the best JS of 0.8284 and the best ASD of 1.14 mm.

The region terms of the active contour model alone, as instantiated by AC, are not sufficient. On spleen MRI with U-Net, AC falls below the plain cross-entropy baseline, at 0.8019 against 0.8139, and on liver CT, where all losses are near saturation, it improves the baseline by no more than 0.003 in DSC. The region term therefore needs to be combined with an explicit boundary regulariser to be reliably useful. Of the two boundary regularisers, the mean curvature formulation of DACMC is consistently more effective than the elastica energy of ACE on the two datasets where the boundary is genuinely ambiguous, particularly in the distance-based metrics. The two losses are equivalent on liver CT, where the boundary is unambiguous. This ordering is consistent with the theoretical distinction between the two terms. The elastica energy penalises the squared curvature, which is dominated by a few high-curvature pixels, whereas the mean curvature term of Eq. (10) averages the curvature over the whole boundary and therefore acts more uniformly along it.

A comparison with prior active-contour and curvature-based losses clarifies the boundary of the present contribution. Four aspects are relevant: which curvature operator is used, how it is computed, how it is coupled with the region term, and what empirical gain has been demonstrated. ACE [31] replaces the length term of the classical active contour model with the elastica energy, so its geometric term is the squared curvature weighted by a total-variation measure; the curvature is obtained from central finite differences of the prediction, that is, by computing first and second derivatives explicitly, and the resulting term is added to the same Chan-Vese region term that DACMC uses. GLAC-UNet [22] couples a global and a local active contour functional for multiclass medical image segmentation, in which the contour evolves according to the learned features rather than through an explicit curvature operator. CDAC [23] conditions the active contour on contextual features extracted by the segmentation network, which modifies the region term rather than the geometric one. Curvature-aware regularisers that act on the network representation, rather than on the geometry of the predicted boundary, penalise or perturb high-curvature activations and therefore constrain the encoded features instead of the segmentation contour itself.

DACMC differs from these formulations in the combination of three properties. It uses an explicit curvature operator rather than a learned or implicit one; it evaluates that operator by convolution, so no derivative of the prediction has to be computed explicitly; and it couples the resulting term to the same Chan-Vese region functional that ACE uses. No prior formulation in this group provides all three at once. The filter and the term that it defines are characterised numerically in Section 6.5, and the empirical consequence is quantified in Tables 1 and 2. Relative to ACE, the closest competing formulation, DACMC reduces HD95 by 1.73 mm and ASD by 1.84 mm on spleen MRI with CE-Net, and reduces HD95 by 3.36 mm on ISIC with CE-Net, whereas the two losses are equivalent on liver CT, where the boundary is unambiguous.

# 6.4. Sensitivity to the curvature weight

Table 3 reports the test DSC obtained by DACMC for $\lambda \in \{10^{-4}, 5\times10^{-4}, 10^{-3}, 5\times10^{-3}\}$ on the three datasets and the two backbones, and Table 4 gives the full five-point sweep on liver CT with CE-Net. For $\lambda \leq 10^{-3}$ the method is well behaved over the entire range. On liver CT the five values in $[0, 10^{-3}]$ give DSC between 0.9776 and 0.9799, and on spleen MRI and ISIC the DSC changes by at most 0.043 as $\lambda$ is increased from $10^{-4}$ to $10^{-3}$. This variation is monotonic in three of the four configurations, and no configuration shows any sign of instability. At $\lambda = 5\times10^{-3}$ the balance of the loss is broken. With CE-Net on spleen MRI the curvature term dominates the region terms to such an extent that the network converges to the trivial all-background solution, producing empty masks on all 25 test slices, with DSC = 0 and undefined HD95 and ASD. On spleen MRI with U-Net the same value degrades the DSC from 0.8348 to 0.7968. This behaviour follows directly from the geometry of the loss. The curvature term of Eq. (10) is minimised by a constant zero field, so an excessively large λ makes the empty prediction an attractor of the optimisation, whereas $\lambda \leq 10^{-3}$ leaves the region terms in control. The lower end of the range is equally benign. With λ = 0, that is with the region terms alone, the liver DSC is 0.9776, essentially identical to the values obtained for small positive λ. The admissible interval for λ is therefore wide, and $\lambda = 10^{-3}$ sits inside the plateau while remaining well below the collapse threshold, so that a single value can be shared by datasets of very different difficulty. Figure 7 illustrates this failure mode on three test slices: with the seed, the backbone and the test cases held fixed, the predictions that are accurate at $\lambda = 1\times10^{-3}$ fragment or erode at $\lambda = 5\times10^{-3}$.

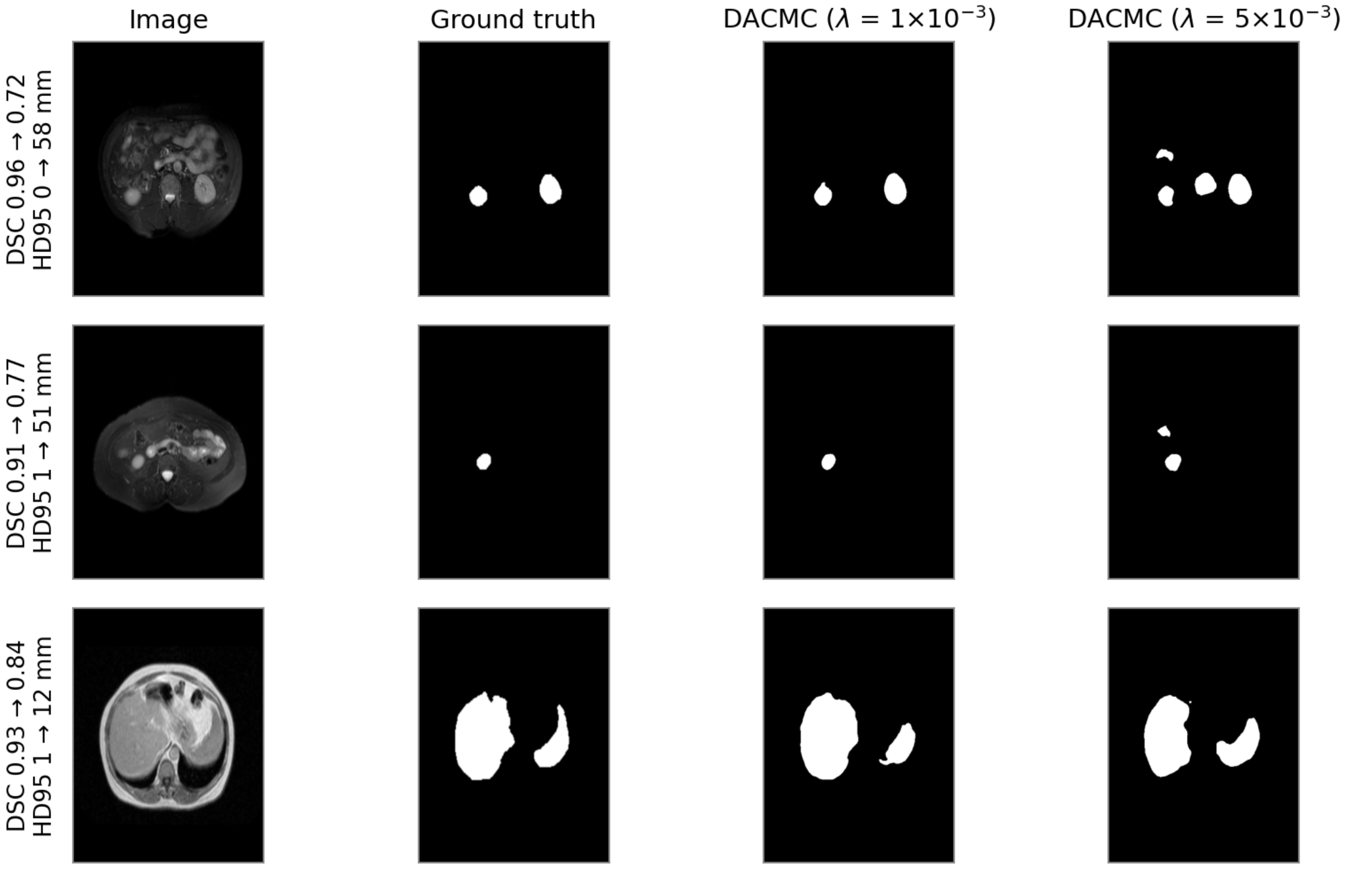

Figure 7. Failure mode of an excessive curvature weight. Identical seed (42), backbone (U-Net) and test cases; only λ changes, from $1\times10^{-3}$ (third column) to $5\times10^{-3}$ (fourth column). Rows show three test slices of the spleen MRI dataset, with the DSC and HD95 of the two predictions given at the left of each row. The predictions that are accurate at $\lambda = 1\times10^{-3}$ fragment or erode at $\lambda = 5\times10^{-3}$, which is the same mechanism that produces the empty masks of Table 3 with CE-Net.

6.5. Cost of the curvature term and accuracy of the filter

**Table 5. Wall-clock training time of the losses under the identical protocol (300 epochs, NVIDIA RTX 5000).**

| Dataset / Backbone | CE | ACE | DACMC ($\lambda = 10^{-3}$) | DACMC vs CE |
|---|---|---|---|---|
| Liver CT / U-Net | 15.2 | 15.4 | **15.7** | +3.3% |
| Liver CT / CE-Net | 12.1 | 12.9 | **12.6** | +4.1% |
| Spleen MRI / U-Net | 24.8 | 25.1 | **25.8** | +4.0% |
| Spleen MRI / CE-Net | 18.9 | 20.2 | **19.8** | +4.8% |
| ISIC dermoscopy / U-Net | 70.8 | 71.7 | **73.5** | +3.8% |
| ISIC dermoscopy / CE-Net | 49.0 | 56.4 | **53.4** | +9.0% |

Values are minutes. DACMC uses $\lambda = 10^{-3}$. The curvature term adds no trainable parameters, so all five losses share the same network and the same inference cost.

Table 5 reports the wall-clock training time of the losses under the identical protocol. Because the curvature term enters only the training objective, it adds no trainable parameters and no computation to the forward pass, so the network topology and the inference cost are identical to those of the baselines. Relative to the cross-entropy baseline, DACMC increases the training time by 4.8% on average, between 3.3% and 9.0% across the six configurations. Relative to the elastica-based ACE loss, the closest competing formulation, DACMC is 0.5% faster on average and faster in three of the six configurations, because the elastica energy requires repeated first- and second-order difference operators whereas DACMC requires a single convolution.

The filter itself is a fourth-order-accurate discrete Laplacian. Its output satisfies $K * u = (3/16)\nabla^2 u$, with a relative residual below $2 \times 10^{-3}$ on smooth fields, so the operator is well defined and introduces no singularity. Its relation to the analytic mean curvature $H$ of a surface $z = u(x, y)$ has a closed form. With the scale that matches the zero-gradient limit, $(8/3)K * u = \nabla^2 u/2$, the relative deviation from $H$ is $|(1 + |\nabla u|^2)^{3/2} - 1|$. We verified this on synthetic surfaces with known curvature: the measured and the predicted deviations agree with a correlation of 0.940 for gradients spanning 0 to 2, and the deviation vanishes as the gradient tends to zero, where the filter and the mean curvature coincide exactly. The deviation is therefore a scale factor that depends on the local gradient rather than an error in the shape of the response.

This scale factor does not propagate to the loss, because the curvature term of Eq. (10) is a normalised ratio between the response of the prediction and the response of the ground truth. Rescaling the kernel by a factor of 1000 changes the value of the term on boundary pixels by 0.15% only. What the term contributes to the optimisation is therefore the spatial agreement of the curvature response, not its absolute magnitude. The response is strictly local: on binary masks it is confined to within one pixel of the true boundary, with a hit rate of 100% and no leakage into the interior of the object or into the background, so the term does not perturb the region terms. The same normalisation also determines the magnitude of λ. On pixels in the interior of the foreground the term is about two orders of magnitude larger than the region terms, and $\lambda = 10^{-3}$ is the value that brings the two to the same order, which is why a larger value makes the curvature term dominate and collapse the prediction to an empty mask (Table 3).

## 6.6. Multi-seed replication and stability

The comparison between DACMC and ACE in Tables 1 and 2 rests on a single training run per configuration. To quantify how much of the reported difference survives run-to-run variability, the two losses were retrained on spleen MRI and ISIC dermoscopy with both backbones over five random seeds (42 to 46), keeping every other setting identical to the protocol of Section 5.3: the same splits, the same optimiser, learning rate, batch size and epoch budget, no data augmentation and no post-processing. This adds 40 training runs. Table 6 reports the mean and the standard deviation of the four metrics over the five seeds, and Table 7 reports paired significance tests between the two losses.

**Table 6.** Mean ± standard deviation of DACMC and ACE over five random seeds (42–46), on spleen MRI and ISIC dermoscopy with both backbones. DACMC uses $\lambda = 10^{-3}$ and ACE uses $\alpha = 10^{-4}$ and $\beta = 1$, as in Tables 1 and 2.

| Dataset / backbone | Loss | DSC ↑ | HD95 ↓ | JS ↑ | ASD ↓ |
|---|---|---|---|---|---|
| Spleen MRI / U-Net | DACMC | **0.823 ± 0.008** | **23.3 ± 6.7** | **0.721 ± 0.014** | **4.32 ± 2.16** |
| Spleen MRI / U-Net | ACE | 0.807 ± 0.023 | 26.8 ± 6.0 | 0.703 ± 0.024 | 7.53 ± 2.63 |
| Spleen MRI / CE-Net | DACMC | **0.807 ± 0.013** | **21.7 ± 5.2** | **0.703 ± 0.018** | 5.20 ± 2.48 |
| Spleen MRI / CE-Net | ACE | 0.791 ± 0.012 | 22.5 ± 2.5 | 0.678 ± 0.013 | **5.07 ± 0.40** |
| ISIC dermoscopy / U-Net | DACMC | 0.864 ± 0.008 | 16.1 ± 2.1 | 0.785 ± 0.011 | 3.19 ± 1.68 |
| ISIC dermoscopy / U-Net | ACE | **0.865 ± 0.006** | **13.7 ± 1.5** | **0.788 ± 0.009** | **2.76 ± 0.40** |
| ISIC dermoscopy / CE-Net | DACMC | 0.876 ± 0.010 | 11.8 ± 1.6 | 0.803 ± 0.011 | 2.74 ± 0.88 |
| ISIC dermoscopy / CE-Net | ACE | **0.877 ± 0.015** | **11.5 ± 2.4** | **0.805 ± 0.014** | **2.44 ± 0.70** |

On spleen MRI the mean over the five seeds favours DACMC in all four metrics with U-Net, and in three of the four metrics with CE-Net: the mean DSC is higher by 0.016 with both backbones, the mean HD95 is lower by 3.5 mm and 0.8 mm, and the mean ASD is lower by 3.2 mm with U-Net and higher by 0.1 mm with CE-Net. On ISIC dermoscopy the two losses are equivalent: the DSC differs by at most 0.002, and ACE attains the lower HD95 by 2.4 mm with U-Net.

**Table 7.** Paired tests between DACMC and ACE. Per-case Wilcoxon signed-rank tests pair the two losses under the same seed and average over seeds (n = 25 on spleen MRI and n = 98 on ISIC dermoscopy); seed-level paired t-tests use the five per-seed means (n = 5). The means are computed over per-case values averaged across seeds and therefore differ marginally from those of Table 6.

| Dataset / backbone | Metric | DACMC | ACE | Δ | Wilcoxon p (per-case) | Paired t p (seed-level) |
|---|---|---|---|---|---|---|
| Spleen MRI / U-Net | DSC ↑ | 0.823 | 0.807 | +0.015 | 0.578 | 0.247 |
| Spleen MRI / U-Net | HD95 ↓ | 23.85 | 26.76 | −2.91 | 0.958 | 0.504 |
| Spleen MRI / CE-Net | DSC ↑ | 0.807 | 0.791 | +0.016 | 0.101 | 0.161 |
| Spleen MRI / CE-Net | HD95 ↓ | 21.94 | 22.52 | −0.58 | 0.542 | 0.667 |
| ISIC dermoscopy / U-Net | DSC ↑ | 0.864 | 0.865 | −0.002 | 0.778 | 0.642 |
| ISIC dermoscopy / U-Net | HD95 ↓ | 16.51 | 14.26 | +2.25 | 0.517 | 0.208 |
| ISIC dermoscopy / CE-Net | DSC ↑ | 0.876 | 0.877 | −0.001 | 0.945 | 0.854 |
| ISIC dermoscopy / CE-Net | HD95 ↓ | 12.03 | 11.72 | +0.31 | 0.516 | 0.795 |

None of the differences in Table 7 reaches the 0.05 level under either test; the smallest p-value is 0.10, for the DSC on spleen MRI with CE-Net. This is consistent with the effect sizes involved: the gaps between the two losses in Tables 1 and 2 are of the order of 0.005 to 0.02 in DSC, and five seeds do not suffice to confirm differences of this size individually. The replication therefore does not establish a significant mean advantage, and we do not claim one.

What the replication does establish is stability. The cross-seed standard deviation of the DSC on spleen MRI with U-Net is 0.008 for DACMC against 0.023 for ACE, and DACMC attains the higher mean DSC in four of the five seeds on both spleen configurations. The worst ACE seed degrades to 0.773 DSC and 11.5 mm ASD, whereas the worst DACMC seed stays at 0.824 DSC and 1.4 mm. The curvature term does not only raise the mean; it removes the occasional severe failure of the elastica term.

One caveat concerns the hardware. The replication was executed on a different GPU than the runs of Tables 1 and 2, an NVIDIA GeForce RTX 4080 SUPER instead of the NVIDIA RTX 5000 used throughout the paper. With U-Net the results reproduce the published values exactly: the seed-42 runs attain 0.8348 and 0.8303 DSC on spleen MRI, and the run at $\lambda = 5\times10^{-3}$ attains 0.7968 DSC, all three identical to the values of Tables 1 and 3. With CE-Net the absolute values shift slightly, because the cuDNN library selects different convolution algorithms for the dilated convolutions of CE-Net on the other architecture; the relative ordering of the two losses is preserved. Every comparison reported in this subsection is paired within this single environment, so the paired tests are unaffected, and the values of Tables 1 and 2 remain those of the original runs.

# 7. Discussion

This section discusses the limitations of the present study, the extent to which the reported comparisons support the conclusions, and the directions in which the method could be extended.

Several limitations should be kept in mind when interpreting these results. First, the three datasets are partitioned in different ways, and this affects how the numbers should be compared. For spleen MRI the 25 test slices come from 4 subjects that are absent from the training partition, so the evaluation is subject-disjoint; for ISIC the test images are distinct images that are never used for training. The liver CT dataset, in contrast, is derived from a single volumetric scan and was partitioned slice by slice, so neighbouring slices, whose boundaries are highly correlated, appear on both sides of the split. The absolute liver DSC values are therefore optimistic and should not be compared with studies that adopt subject-level splits. What remains valid is the relative comparison between losses, because all five are trained and evaluated on exactly the same split.

Second, the validation subsets used for model selection and for the choice of λ are drawn from the same partition as the training data. Selecting a hyper-parameter on such a subset can be optimistic. We mitigate this in two ways: a single $\lambda = 10^{-3}$ is shared by all six settings rather than tuned per setting, and Table 3 shows that the performance is insensitive to λ over the whole range $\lambda \leq 10^{-3}$ that was tested, so the reported results are not driven by validation noise. Nevertheless, a protocol in which the validation subset is subject-disjoint from the training data would be preferable and is left to future work.

Third, the ISIC test set contains 98 rather than 100 images. An exhaustive byte-level and perceptual-hash scan of the 500-image pool revealed two pairs in which a training image and a test image are identical while their masks disagree; such duplicates are a known artefact of the ISIC archive. Because the test-side masks could not be trusted, the two test images were removed and the training portion was left unchanged.

Fourth, each configuration was trained once over a single random split, without data augmentation, and the loss functions were compared at a single learning rate. The spread across datasets is much larger than the differences between losses on liver CT, so differences of the order of 0.002 DSC on that dataset should be treated as within run-to-run variability. A five-seed replication of the comparison between DACMC and ACE on spleen MRI and ISIC dermoscopy is reported in Section 6.6; it shows that the mean advantage of DACMC does not reach statistical significance at five seeds, while the curvature term removes the occasional severe failure of the elastica term. Extending the replication to all five losses and to liver CT with subject-level splits, and moving to 3-D volumetric models, is left for future work.

Finally, the convolution-kernel filter yields a local, single-scale curvature response, which may not fully capture curvature variations across multiple scales or preserve directional curvature information on structures with sharp corners or thin branches. The loss has likewise been validated only on 2-D single-organ datasets. Future directions include integrating DACMC with transformer-based segmentation architectures, developing learnable or adaptive curvature operators, and extending the framework to 3-D multi-organ and multi-modal volumetric medical images.

# 8. Conclusion

This paper proposed a Deep Active Contour and Mean Curvature (DACMC) loss function for medical image segmentation that addresses the lack of geometric prior information inherent to pixel-level supervised learning. The mean curvature is incorporated into the Chan-Vese region functional, the two-phase piecewise-constant form of the Mumford-Shah energy, providing a second-order geometric regulariser that characterises the segmented region beyond the conventional region and length terms. To circumvent the high-order derivatives and non-smoothness of the variational mean curvature, a fixed convolution-kernel mean curvature filter is employed, which preserves differentiability for back-propagation, adds no trainable parameters, and increases the training time by 4.8% relative to the cross-entropy baseline while leaving the inference cost unchanged. The resulting loss couples the region-based discriminability inherited from the Chan-Vese/Mumford-Shah model with boundary regularity through curvature, and requires only one weighting parameter $\lambda$ to be tuned.

The loss was evaluated on three public datasets – liver CT, spleen MRI and ISIC dermoscopy – with the U-Net and CE-Net backbones and four metrics, DSC, HD95, JS and ASD, against the cross-entropy, Dice, active contour and elastica losses trained under an identical protocol. The results support a precise claim. DACMC is the best or the second-best method in DSC in five of the six dataset-backbone settings, and its most consistent gains are in the distance-based metrics, where it ranks first or second in 10 of 12 combinations. On spleen MRI with CE-Net it reduces HD95 to 16.28 mm and ASD to 1.90 mm, and on ISIC with CE-Net it reduces HD95 to 7.08 mm, 29% below the next best method. On liver CT, where all losses are close to saturation, the five methods are statistically indistinguishable. This pattern is consistent with the role of the curvature term as a boundary regulariser, whose contribution is largest when the target is small and varies in scale. A sensitivity study further clarifies the behaviour of the single hyper-parameter of the method.

For $\lambda \leq 10^{-3}$ the DSC varies by at most 0.0023 on liver CT over three orders of magnitude and by at most 0.043 on the two harder datasets, whereas $\lambda = 5\times10^{-3}$ causes the curvature term to dominate the loss and the network to collapse to an empty prediction on spleen MRI. The admissible range of $\lambda$ is therefore wide, which makes the method straightforward to deploy. A five-seed replication against the elastica loss further indicates that the curvature term improves the stability of the training outcome, although the mean differences do not reach statistical significance at five seeds.

## CRediT authorship contribution statement

Xiao-qiang Zhai: Conceptualization, Methodology, Software, Formal analysis, Investigation, Writing – original draft. Zhi-feng Pang: Supervision, Writing – review & editing. Peng Zheng: Software, Data curation. Ze-wen Li: Validation, Visualization. Yan-zhe Hou: Resources, Writing – review & editing. All authors read and approved the final manuscript.

## Funding

This research received no external funding.

## Ethics statement

Not applicable. This study used only publicly available, de-identified imaging datasets and involved no human participants recruited by the authors; institutional ethical approval and informed consent were therefore not required.

## Data availability

The datasets used in this study are publicly available. The liver CT images are available at https://www.kaggle.com/datasets/zxcv2022/digital-medical-images-for--download-resource, the spleen MRI images at https://chaos.grand-challenge.org/ and the dermoscopy images through the ISIC archive (https://www.isic-archive.com/). For the dermoscopy experiments we used a publicly available curated subset of 500 images with binary lesion masks, partitioned into 400 training/validation and 98 test images, as described in Section 5.1.

Code availability: the implementation of the DACMC loss and of the discrete mean-curvature filter is fully specified by Eq. (15), the kernel coefficients reported in Section 4 and the training protocol of Section 5.3. The implementation is available from the corresponding author upon reasonable request.

## Declaration of generative AI use

During the preparation of this work the authors used AI-assisted tools for language polishing, formatting, and organisational editing of the manuscript. After using these tools, the authors reviewed and edited the content as needed and take full responsibility for the content of the publication.

## Declaration of Competing Interest

The authors declare no conflicts of interest.

# References

**[1]** Litjens G, Kooi T, Bejnordi BE, Setio AAA, Ciompi F, Ghafoorian M, van der Laak JAWM, van Ginneken B, Sanchez CI. A survey on deep learning in medical image analysis. Med. Image Anal. 2017;42:60-88. doi:10.1016/j.media.2017.07.005

**[2]** Ronneberger O, Fischer P, Brox T. U-Net: convolutional networks for biomedical image segmentation. In: Medical Image Computing and Computer-Assisted Intervention - MICCAI 2015. Lect. Notes Comput. Sci. 2015;9351:234-241. doi:10.1007/978-3-319-24574-4_28

**[3]** Badrinarayanan V, Kendall A, Cipolla R. SegNet: a deep convolutional encoder-decoder architecture for image segmentation. IEEE Trans. Pattern Anal. Mach. Intell. 2017;39(12):2481-2495. doi:10.1109/TPAMI.2016.2644615

**[4]** Simonyan K, Zisserman A. Very deep convolutional networks for large-scale image recognition. arXiv:1409.1556. 2015. Preprint.

**[5]** Szegedy C, Liu W, Jia Y, Sermanet P, Reed S, Anguelov D, Erhan D, Vanhoucke V, Rabinovich A. Going deeper with convolutions. In: Proceedings of the IEEE Conference on Computer Vision and Pattern Recognition (CVPR). 2015. p. 1-9. doi:10.1109/CVPR.2015.7298594

**[6]** Singh A, Sengupta S, Lakshminarayanan V. Explainable deep learning models in medical image analysis. J. Imaging. 2020;6(6):52. doi:10.3390/jimaging6060052

**[7]** Tajbakhsh N, Jeyaseelan L, Li Q, Chiang JN, Wu Z, Ding X. Embracing imperfect datasets: a review of deep learning solutions for medical image segmentation. Med. Image Anal. 2020;63:101693. doi:10.1016/j.media.2020.101693

**[8]** Zhang J, Guo W. A new regularization for deep learning-based segmentation of images with fine structures and low contrast. Sensors. 2023;23(4):1887. doi:10.3390/s23041887

**[9]** Tan L, Ma S, Xu J, Pan J, Yuan J, Kervadec H, Pelillo M. Spatially continuous dual optimization on compactness function for image segmentation. Pattern Recognit. 2026;172:112613. doi:10.1016/j.patcog.2025.112613

**[10]** Long J, Shelhamer E, Darrell T. Fully convolutional networks for semantic segmentation. In: Proceedings of the IEEE Conference on Computer Vision and Pattern Recognition (CVPR). 2015. p. 3431-3440. doi:10.1109/CVPR.2015.7298965

**[11]** Sudre CH, Li W, Vercauteren T, Ourselin S, Cardoso MJ. Generalised Dice overlap as a deep learning loss function for highly unbalanced segmentations. In: Deep Learning in Medical Image Analysis and Multimodal Learning for Clinical Decision Support. Lect. Notes Comput. Sci. 2017;10553:240-248. doi:10.1007/978-3-319-67558-9_28

**[12]** Condat L, Kitahara D, Contreras A, Hirabayashi A. Proximal splitting algorithms for convex optimization: a tour of recent advances, with new twists. SIAM Rev. 2023;65(2):375-435. doi:10.1137/20M1379344

**[13]** Terven J, Cordova-Esparza DM, Romero-Gonzalez JA, Ramirez-Pedraza A, Chavez-Urbiola EA, et al. A comprehensive survey of loss functions and metrics in deep learning. Artif. Intell. Rev. 2025;58(7):195. doi:10.1007/s10462-025-11198-7

**[14]** Zhang K, Li L, Liu H, Yuan J, Tai XC. Deep convolutional neural networks meet variational shape compactness priors for image segmentation. Neurocomputing. 2025;623:129395. doi:10.1016/j.neucom.2025.129395

**[15]** Wen B, Zhang H, Bartsch DG, et al. Topology-preserving image segmentation with spatial-aware persistent feature matching. arXiv:2412.02076. 2025. Preprint.

**[16]** Kervadec H, Bouchtiba J, Desrosiers C, Granger E, Dolz J, Ben Ayed I. Boundary loss for highly unbalanced segmentation. Med. Image Anal. 2021;67:101851. doi:10.1016/j.media.2020.101851

**[17]** Zhang S, Jia F, Li X, Zhang H, Shi J, Ma L, Ying S. LMS-Net: a learned Mumford-Shah network for binary few-shot medical image segmentation. Med. Image Anal. 2025;105:103676. doi:10.1016/j.media.2025.103676

**[18]** Chen X, Williams BM, Vallabhaneni SR, Czanner G, Williams R, Zheng Y. Learning active contour models for medical image segmentation. In: Proceedings of the IEEE/CVF Conference on Computer Vision and Pattern Recognition (CVPR). 2019. p. 11632-11640. doi:10.1109/CVPR.2019.01190

**[19]** Qi K, Huang Z, Yang W. Robust variational model based tailored UNet: leveraging edge detector and mean curvature for improved image segmentation. arXiv:2512.07590. 2025. Preprint.

**[20]** Hajrasooliha M, Naghsh-Nilchi AR. A hybrid framework integrating active contour and deep learning for optic disc and optic cup segmentation. Signal Image Video Process. 2026;20(6):350. doi:10.1007/s11760-026-05378-3

**[21]** Lambert Z, Petitjean C, Le Guyader C. Blending variational approaches and deep learning to enforce prior constraints in medical image segmentation. In: Less-Supervised Segmentation with CNNs. Elsevier; 2026. p. 217-235. doi:10.1016/B978-0-32-395674-1.00017-2

**[22]** Trinh MN, Tran TT, Nham DHN, Lo MT, Pham VT. GLAC-UNet: global-local active contour loss with an efficient U-shaped architecture for multiclass medical image segmentation. J. Imaging Inform. Med. 2025;38(5):3198-3220. doi:10.1007/s10278-025-01387-9

**[23]** da Silva SPP, Ivo RF, Barroso CB, Fernandes JCN, Portela TF, Medeiros AG, de Sousa PHF, Song H, Reboucas Filho PP. Context-driven active contour (CDAC): a novel medical image segmentation method based on active contour and contextual understanding. Sensors. 2025;25(9):2864. doi:10.3390/s25092864

**[24]** Li W, Tai XC, Liu J. Topology-guaranteed image segmentation: enforcing connectivity, genus, and width constraints. SIAM J. Imaging Sci. 2026;19(2):1137-1173. doi:10.1137/25M1765870

**[25]** Milletari F, Navab N, Ahmadi S. V-Net: fully convolutional neural networks for volumetric medical image segmentation. In: 2016 Fourth International Conference on 3D Vision (3DV). IEEE; 2016. p. 565-571. doi:10.1109/3DV.2016.79

[26] Mumford D, Shah J. Optimal approximations by piecewise smooth functions and associated variational problems. Commun. Pure Appl. Math. 1989;42(5):577-685. doi:10.1002/cpa.3160420503

**[27]** Tai XC, Li L, Bae E. The Potts model with different piecewise constant representations and fast algorithms: a survey. In: Handbook of Mathematical Models and Algorithms in Computer Vision and Imaging. Springer; 2023. p. 1887-1927. doi:10.1007/978-3-030-98661-2_90

**[28]** Chan TF, Vese LA. Active contours without edges. IEEE Trans. Image Process. 2001;10(2):266-277. doi:10.1109/83.902291

**[29]** Kim B, Ye JC. Mumford-Shah Loss Functional for Image Segmentation with Deep Learning. IEEE Trans. Image Process. 2020;29:1856-1866. doi:10.1109/TIP.2019.2941265

**[30]** Mumford D. Elastica and computer vision. In: Bajaj C, editor. Algebraic Geometry and its Applications. Springer; 1994. p. 491-506.

**[31]** Chen X, Luo X, Wang G, Zheng Y. Deep Elastica for image segmentation. In: 2021 IEEE 18th International Symposium on Biomedical Imaging (ISBI). IEEE; 2021. p. 706-710. doi:10.1109/ISBI48211.2021.9433886

**[32]** Zhu W, Chan T. Image denoising using mean curvature of image surface. SIAM J. Imaging Sci. 2012;5(1):1-32. doi:10.1137/110822268

**[33]** do Carmo MPL. Differential Geometry of Curves and Surfaces. Englewood Cliffs, NJ: Prentice-Hall; 1976.

**[34]** Dong G, Hintermuller M, Zhang Y. A class of second-order geometric quasilinear hyperbolic PDEs and their application in imaging. SIAM J. Imaging Sci. 2021;14(2):645-688. doi:10.1137/20M1366277

**[35]** Rudin LI, Osher S, Fatemi E. Nonlinear total variation based noise removal algorithms. Physica D. 1992;60(1-4):259-268. doi:10.1016/0167-2789(92)90242-F

**[36]** Zhu W, Tai XC, Chan T. Augmented Lagrangian method for a mean curvature based image denoising model. Inverse Probl. Imaging. 2013;7(4):1409-1432. doi:10.3934/ipi.2013.7.1409

**[37]** Xing G, Chen L, Wang H, Zhang J, Sun D, Xu F, Lei J, Xu X. Multi-scale pathological fluid segmentation in OCT with a novel curvature loss in convolutional neural network. IEEE Trans. Med. Imaging. 2022;41(6):1547-1559. doi:10.1109/TMI.2022.3142048

**[38]** Gong Y, Sbalzarini IF. Curvature filters efficiently reduce certain variational energies. IEEE Trans. Image Process. 2017;26(4):1786-1798. doi:10.1109/TIP.2017.2658954

**[39]** Kavur AE, Gezer NS, Baris M, Aslan S, Conze PH, et al. CHAOS challenge - combined (CT-MR) healthy abdominal organ segmentation. Med. Image Anal. 2021;69:101950. doi:10.1016/j.media.2020.101950

**[40]** Gu Z, Cheng J, Fu H, Zhou K, Hao H, Zhao Y, Zhang T, Gao S, Liao J, et al. CE-Net: context encoder network for 2D medical image segmentation. IEEE Trans. Med. Imaging. 2019;38(10):2281-2292. doi:10.1109/TMI.2019.2903562